\documentclass{article}

\usepackage{arxiv}

\usepackage[T1]{fontenc}    
\usepackage{url}            
\usepackage[ toc, page, title, titletoc ]{appendix}
\usepackage[ruled,noline]{algorithm2e}
\usepackage{amsmath}
\usepackage{amsfonts}
\usepackage{booktabs}
\usepackage[colorinlistoftodos]{todonotes}
\usepackage{hyperref}
\usepackage{lineno}
\usepackage{mathrsfs}
\usepackage{float}
\usepackage{url}            

\usepackage{etoolbox}
\providetoggle{fullBackground}

\usepackage{parskip}  

\DeclareFontFamily{U}{wncy}{}
\DeclareFontShape{U}{wncy}{m}{n}{<->wncyr10}{}
\DeclareSymbolFont{mcy}{U}{wncy}{m}{n}
\DeclareMathSymbol{\comb}{\mathord}{mcy}{"58} 

\usepackage{accents}
\newlength{\dhatheight}

\title{Accelerated Parallel Magnetic Resonance Imaging with Compressed Sensing using Structured Sparsity}

\author{
  Nicholas Dwork\thanks{www.nicholasdwork.com, nicholas.dwork@cuanschutz.edu} \\
  Departments of Pediatrics and Radiology \\
  University of Colorado | Anschutz Medical Center
    \And
  Erin K. Englund \\
  Department of Radiology \\
  University of Colorado | Anschutz Medical Center
}

\begin{document}
\maketitle

\begin{abstract}
  Compressed sensing is an imaging paradigm that allows one to invert an underdetermined linear system by imposing the \textit{a priori} knowledge that the sought after solution is sparse (i.e., mostly zeros).  Previous works have shown that if one also knows something about the sparsity pattern (the locations where non-zero entries exist), one can take advantage of this structure to improve the quality of the result.  A significant application of compressed sensing is magnetic resonance imaging (MRI), where samples are acquired in the Fourier domain.  Compressed sensing allows one to reconstruct a high-quality image with fewer samples which can be collected with a faster scan.  This increases the robustness of MRI to patient motion since less motion is possible during the shorter scan.  Parallel imaging, where multiple coils are used to gather data, is another an more ubiquitously used method for accelerating MRI.  Existing combinations of these acceleration methods, such as Sparse SENSE, yield high quality images with an even shorter scan time than either technique alone.  In this work, we show how to modify Sparse SENSE with structured sparsity to reconstruct a high quality image with even fewer samples.
\end{abstract}


\section{Introduction}
\label{sec:intro}

Magnetic resonance imaging (MRI) is a ubiquitously used gross imaging modality due to its ability to image with significant natural contrast (without any exogenous contrast agent) and its complete lack of ionizing radiation.  MRI acquires samples in the frequency domain.  With a fully sampled reconstruction, enough data to satisfy the Nyquist-Shannon sampling theorem is collected, and the image is reconstructed with a simple inverse Fast Fourier Transform (IFFT).  Because of this, MRI requires that the patient remain still during the scan.  This is especially challenging for three-dimensional MRI, which requires scan times up to 10 minutes for conventional reconstruction.  Two methods of accelerating MRI include parallel imaging and compressed sensing.  Parallel imaging uses multiple sensing coils (i.e., antennas) to simultaneously image the subject from different vantage points \cite{sodickson1997simultaneous,deshmane2012parallel}.  As we will show below, the unique information provided by each antenna can be used to interpolate missing Fourier values and reconstruct a high-quality image.  Another method to accelerate MRI is compressed sensing, which relies on the assumption that the image is sparse after an invertible sparsifying transformation.  By combining these methods, MRI requires even fewer samples for a high-quality image, which can be collected with an even faster scan.

Three-dimensional MRI with compressed sensing and parallel imaging still requires approximately 30 seconds of scan time \cite{vasanawala2011practical}.  While this is much faster than the conventional fully-sampled acquisition, any further increase in speed could make MRI even more robust to motion or increase patient throughput.  In previous work, we showed that compressed sensing could be accelerated with structured sparsity \cite{dwork2021utilizing,dwork2022utilizing}.  The sparsifying transformations used with compressed sensing are commonly the wavelet \cite{lustig2008compressed,baron2018rapid} and/or curvelet \cite{ma2010improved} transforms, which benefit from fast implementations \cite{beylkin1991fast,candes2006fast}.  Both the wavelet and curvelet transforms apply a low-pass filter of the image; most natural images, and certainly anatomical MR images, have high energy in the low frequencies.  Thus, one would not expect that the intensities of the coefficients corresponding to these low-frequencies would be sparse.  In \cite{dwork2021utilizing,dwork2022utilizing}, Dwork et al. modified the standard optimization problem solved for a compressed sensing reconstruction to take this into account.  By doing so, they were able to generate images of higher quality for a given number of samples.

In this work, we will combine model-based parallel imaging \cite{fessler2010model} with compressed sensing using structured sparsity.  By doing so, we will show that we can recover high-quality images with MRI using even fewer samples, which can be collected with a faster scan\footnote{An early version of this work was submitted for presentation to the 2024 Symposium of the International Society for Magnetic Resonance in Medicine.}.  Matlab code used for this project are shared at \url{https://github.com/ndwork/picsWithStructuredSparsity.git}.

\section{Methods}
\label{sec:methods}

\subsection{Background}
\label{sec:background}

With parallel MRI, multiple sensing coils are used to simultaneously collect data of a patient.  With model-based reconstruction, it is assumed that the sensitivity of each coil is known, which specifies how well the coil senses from each point in space.  The image is reconstructed by solving the following least-squares problem:
\begin{equation}
  \label{eq:mbr}
  \underset{x}{\text{minimize}} \hspace{0.5em} \left\|\boldsymbol{M} \, \boldsymbol{F} \, \boldsymbol{S} \, x - \boldsymbol{b} \right\|_2,
\end{equation}
where $x$ represents the image, $\|\cdot\|$ represents the $\ell_2$ norm, $\boldsymbol{S}$ is a block-column matrix such that $\boldsymbol{S}=\left( S^{(1)}, S^{(2)}, \ldots, S^{(C)} \right)$, $S^{(i)}$ is a diagonal matrix of complex values that represents the sensitivity map of the $i^{\text{th}}$ coil, $C$ is the number of coils used for data collection, $\boldsymbol{F}=\text{diag}\left(F,F,\ldots,F\right)$ is a block-diagonal matrix that applies the FFT to each $S^{(i)}x$ product, $\boldsymbol{M}=(M,M\ldots,M)$ is a diagonal matrix and $M$ represents the data sampling mask, and $\boldsymbol{b}=\left( b^{(1)}, b^{(2)}, \ldots, b^{(C)} \right)$ is a block-column matrix where $b^{(i)}$ represents the data collected by the $i^{\text{th}}$ coil.
When $\boldsymbol{M}\,\boldsymbol{F}\,\boldsymbol{S}$ is full-rank (either invertible of over-determined), then the image is uniquely estimated by solving this problem.  Problems of this form can be solved with the conjugate gradient method or with LSQR \cite{paige1982lsqr}.

With a Fourier sensing apparatus, like MRI, one reconstructs an image using compressed sensing by solving problems of the form
\begin{equation}
  \label{eq:cs}
  \underset{z}{\text{minimize}} \hspace{0.5em} (1/2)\left\| M \, F \, \Psi^\ast \, z - b \right\|_2^2 + \lambda \, \| z \|_1,
\end{equation}
where $\Psi$ is a vector of the sparsifying transformation, $\Psi^\ast$ is its adjoint, and $\lambda>0$ is a regularization parameter.  Note that $\Psi^\ast$ need not be invertible; it can represent an overcomplete basis (e.g., consisting of the wavelet and curvelet transformations).  When \eqref{eq:cs} satisfies the Restricted Isometry Property in Levels, then its solution solves the corresponding sparse signal recovery problem \cite{donoho2005stable,bastounis2017absence}.  Let $z^\star$ be the solution to \eqref{eq:cs}; then the image is reconstructed with $x^\star = \Psi^\ast\,z^\star$.

Sparse SENSE combines model-based reconstruction with compressed sensing; the image is reconstructed by solving the following optimization problem:
\begin{equation*}
  \label{prob:sparseSENSE}
  \underset{x}{\text{minimize}} \hspace{0.5em} (1/2)\left\|\boldsymbol{M} \, \boldsymbol{F} \, \boldsymbol{S} \, x
    - \boldsymbol{b} \right\|_2^2 + \lambda \|\Psi\,x\|_1.
\end{equation*}
Problems of this form can be solved with the Fast Iterative Shrinkage Threshold Algorithm (FISTA) \cite{beck2009fast}.

\subsection{Model-Based Reconstruction with Compressed Sensing Using Structured Sparsity}
\label{sec:methods}

Rather than work with the analysis form of the optimization problem used with Sparse SENSE, we will use the related synthesis formulation \cite{candes2011compressed, boyer2019compressed}:
\begin{equation}
  \underset{z}{\text{minimize}} \hspace{0.5em} (1/2)\left\| \boldsymbol{M} \, \boldsymbol{F} \, \boldsymbol{S} \, \Psi^\ast \, z - \boldsymbol{b} \right\|_2^2 + \lambda \| z \|_1.
  \label{prob:sparseSenseSynthesis}
\end{equation}

Note that if one had an estimate of the low-pass filtered image, then one could modify this problem accordingly:
\begin{equation}
  \underset{z}{\text{minimize}} \hspace{0.5em} (1/2)\left\| \boldsymbol{M} \, \boldsymbol{F} \, 
      \left( \boldsymbol{S} \, \Psi^\ast z + \boldsymbol{x_L} \right) - \boldsymbol{b} \right\|_2^2 +
    \lambda \| z \|_1,
  \label{prob:picsSynthesis}
\end{equation}
where $\boldsymbol{x_L}$ is a vector of low-frequency estimates (blurry images) of each coil.  Since we do not expect for the low-frequencies of the image to be sparse, we choose to satisfy the Nyquist-Shannon sampling theorem for the low-frequency portion of the image.  The low-frequency, blurry images are estimated with $\boldsymbol{x_L}= \boldsymbol{F}^{-1} \, \boldsymbol{K_B} \, \boldsymbol{M_L} \, \boldsymbol{b} $, where $M_L$ is a block-diagonal matrix of a repeated block with values equal to $1$ or $0$ that isolates the low-frequencies according to the two-level sampling scheme of \cite{adcock2017breaking}, and $\boldsymbol{K_B}$ is a repeated block-diagonal matrix that applies the Kaiser-Bessel window \cite{jackson1991selection} as in \cite{dwork2021utilizing} and \cite{dwork2022utilizing}.  By letting $\boldsymbol{\beta} = \boldsymbol{b} - \boldsymbol{K_B} \, \boldsymbol{M_L} \, \boldsymbol{b}$, problem \eqref{prob:picsSynthesis} becomes
\begin{equation}
  \underset{z}{\text{minimize}} \hspace{0.5em} (1/2)\left\| \boldsymbol{M} \, \boldsymbol{F} \, \boldsymbol{S} \, \Psi^\ast z - \boldsymbol{\beta} \right\|_2^2 +
    \lambda \| z \|_1,
  \label{prob:picsStructuredSparsity}
\end{equation}
Problem \eqref{prob:picsStructuredSparsity} is the novel combination of parallel imaging and compressed sensing with structured sparsity.  Note that this problem is of the same form as that of \eqref{prob:sparseSenseSynthesis} where the only difference is that $\boldsymbol{b}$ has been replaced with $\boldsymbol{\beta}$, so it too can be solved with FISTA.  Let $z^\star$ be the solution to problem \eqref{prob:picsStructuredSparsity}; then the images of all coils are reconstructed according to $\boldsymbol{x}^\star = \boldsymbol{x_L} + \boldsymbol{S} \, \boldsymbol{\Psi}^\ast \, z^\star$.  Once the images of all coils are reconstructed, the final image is reconstructed using the method of Roemer \cite{roemer1990nmr}.

The model-based reconstruction presented in \eqref{prob:picsStructuredSparsity} that combines parallel imaging with compressed sensing using structured sparsity amounts to a three-step process for image reconstruction: 1) estimate the blurry images $\boldsymbol{x_L}$, 2) estimated the missing details by solving \eqref{prob:picsStructuredSparsity}, and 3) combine the reconstructions from all coils into a single image.  It is crucial for this approach that the low-frequency region be fully-sampled.

\section{Results}
\label{sec:results}

\subsection{Experimental Setup}
\label{sec:experiments}

\begin{figure}
    \centering
    \includegraphics[width=0.5\linewidth]{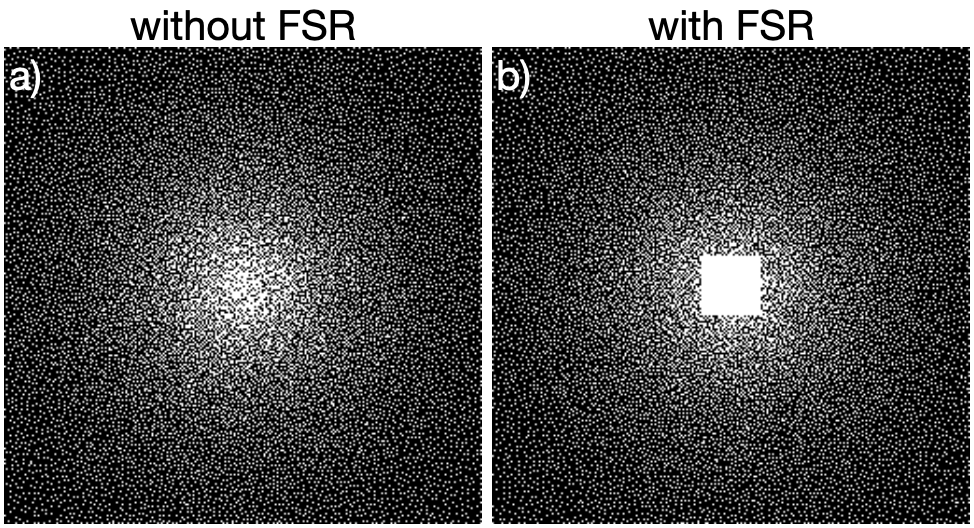}
    \caption{Variable density Poisson disc sampling patterns (left) without and (right) with a fully-sampled region created according to the discrete Daubechies-4 wavelet transform.  Each white point represents a line of data out of the page that was collected.  These sampling patterns create a sampling burden of $24\%$. }
    \label{fig:vdpdPatterns}
\end{figure}

All experiments are from fully-sampled data of anatomies that remain still.  Fully sampled reconstructions were generated by the method of Roemer \cite{roemer1990nmr}.  Results will be shown for data with a brain, knee, ankle, and shoulder.  The fully-sampled reconstruction will be compared to the reconstructions from retrospectively undersampled data.  All data were collected on Cartesian trajectories with two dimensions of phase encodes and one dimension of readout.  The sampling patterns used with be a variable density Poisson disc sampling pattern (without directional variation) created according to \cite{dwork2021fast}; an example is shown in Fig. \ref{fig:vdpdPatterns}.  Unless otherwise stated, the sampling pattern will be augmented with a centered fully-sampled region (FSR).   After inverse Fourier transforming along the readout direction, the data is placed in a $k_x,k_y,z$ hybrid domain where each slice (i.e., individual $z$ locations) can then be processed independently.  We will show results for individual slices from each dataset.  Each problem was solved with values of $\lambda$ equal to $0.001$, $0.002$, $\ldots$, $0.01$, $0.02$, $\ldots$, $0.1$, $0.2$, $\ldots$, $1$, $2$, $\ldots$, $10$.  Unless otherwise stated, the image that achieves the highest Pearson Correlation Coefficient (PCC) value is reported \cite{benesty2008importance}.

\subsection{Results with Retrospective Downsampling}

\begin{figure}
    \centering
    \includegraphics[width=0.7\linewidth]{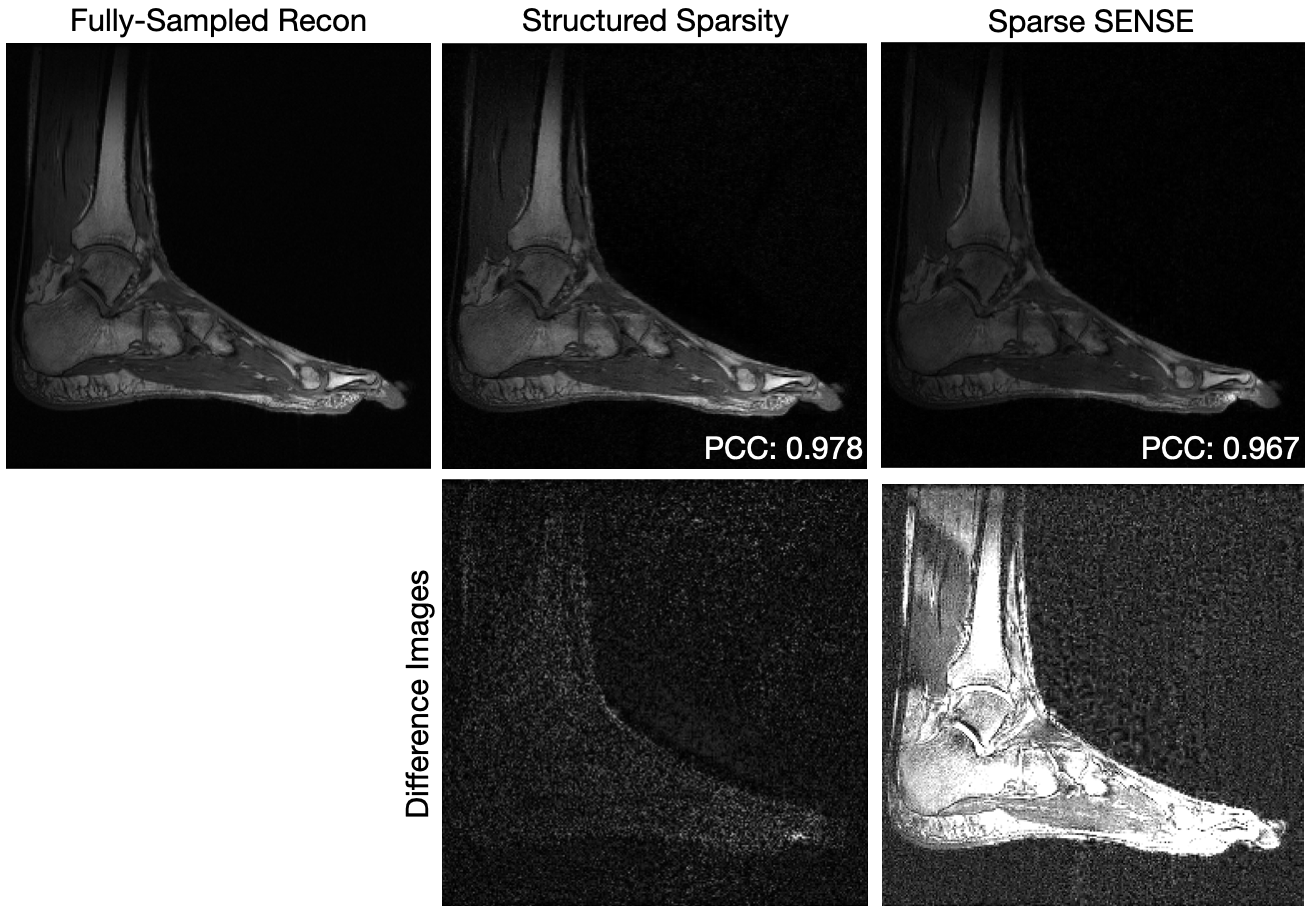}
    \caption{Comparison of fully-sampled reconstruction to reconstructions from accelerated acquisitions for a sagittal slice of an ankle with the compressed sensing with structured sparsity presented in this manuscript and the previously existing Sparse SENSE.  The data collected had an acceleration factor of $4.5$.  Differences with a fully-sampled reconstruction are shown on the same intensity scale.  The Pearson Correlation Coefficient (PCC) is displayed for each reconstruction.  Difference images are shown on the same scale.}
    \label{fig:ankleResult}
\end{figure}

Figure \ref{fig:ankleResult} shows a comparison between the fully-sampled reconstruction, Sparse SENSE, and parallel imaging with compresses sensing using structured sparsity for data of a sagittal slice of an ankle collected with a 8-channel dedicated ankle coil array.  The sampling pattern had an acceleration factor of $4.5$ (i.e., only $22\%$ of the number of samples required to satisfy the Nyquist-Shannon sampling theorem were collected).  For the ankle, the discrete Daubechies-4 wavelet transform was used as the sparsifying transformation \cite{majumdar2012choice}.  The PCC values comparing the undersampled reconstructions and the fully-sampled reconstruction show that structured sparsity (PCC=$0.978$) is more similar to the fully-sampled reconstruction than Sparse SENSE (PCC=$0.967$).  The difference images show that the errors in Sparse SENSE are not isolated to a small region, but instead, are spread throughout the image.  Though the details remain visible with Sparse SENSE, the low-frequencies are highly corrupted.

\begin{figure}
    \centering
    \includegraphics[width=0.7\linewidth]{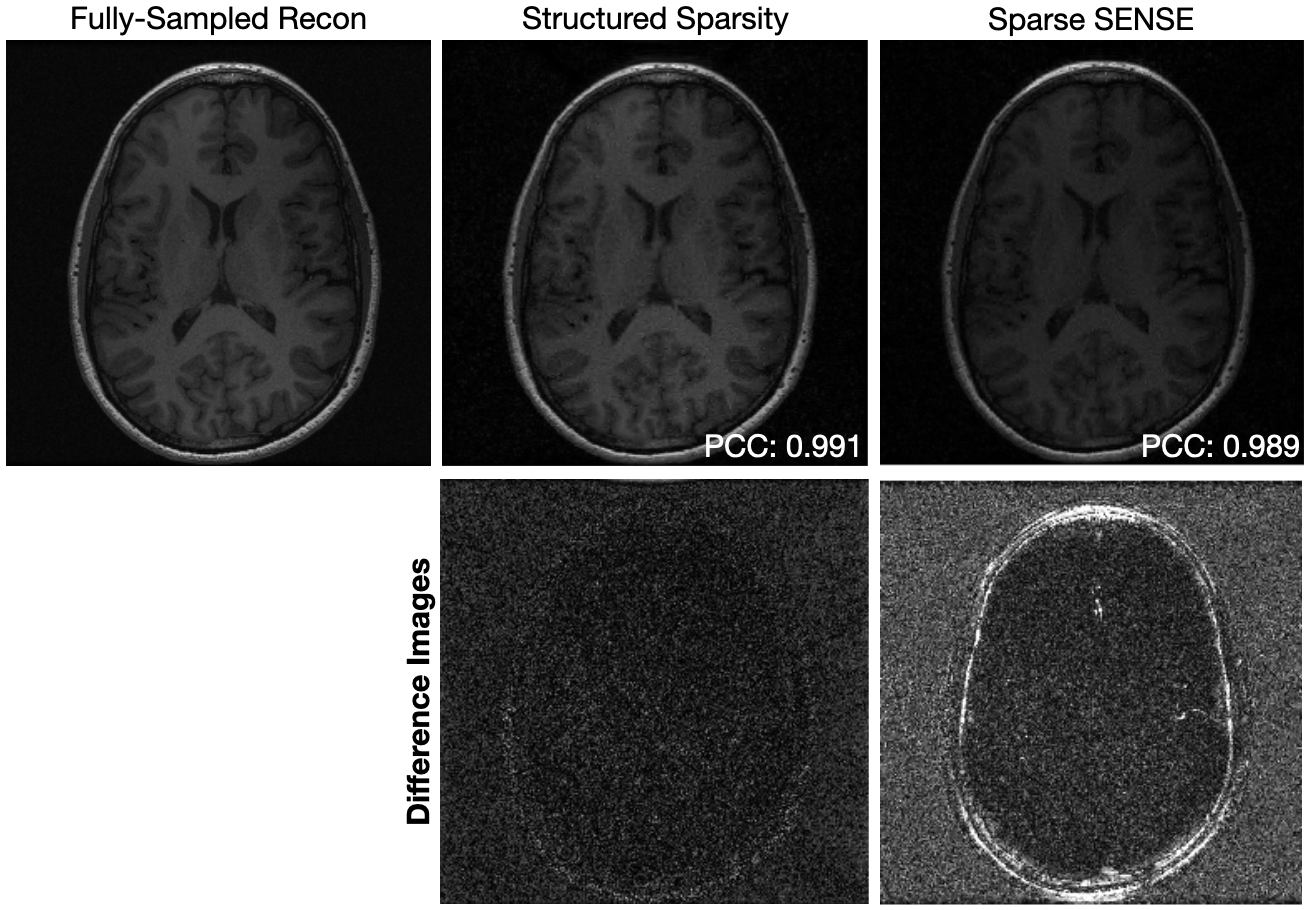}
    \caption{Reconstructions of an axial slice of a brain for compressed sensing with structured sparsity and Sparse SENSE.  The data collected had an acceleration factor of $5$.  Differences with a fully-sampled reconstruction are shown on the same intensity scale.  The Pearson Correlation Coefficient (PCC) is displayed for each reconstruction.  Difference images are shown on the same scale.}
    \label{fig:brainResult}
\end{figure}

Figure \ref{fig:brainResult} shows a similar comparison between the fully-sampled reconstruction, and reconstructions from $20\%$ of the fully-sampled data using Sparse SENSE, and parallel imaging with compressed sensing using structured sparsity for data of an axial slice of a brain collected with an 8-channel birdcage coil.  As with the ankle, compressed sensing with structured sampling performs better than Sparse SENSE (PCC=$0.991$ and PCC=$0.989$, respectively).

\begin{figure}
    \centering
    \includegraphics[width=0.7\linewidth]{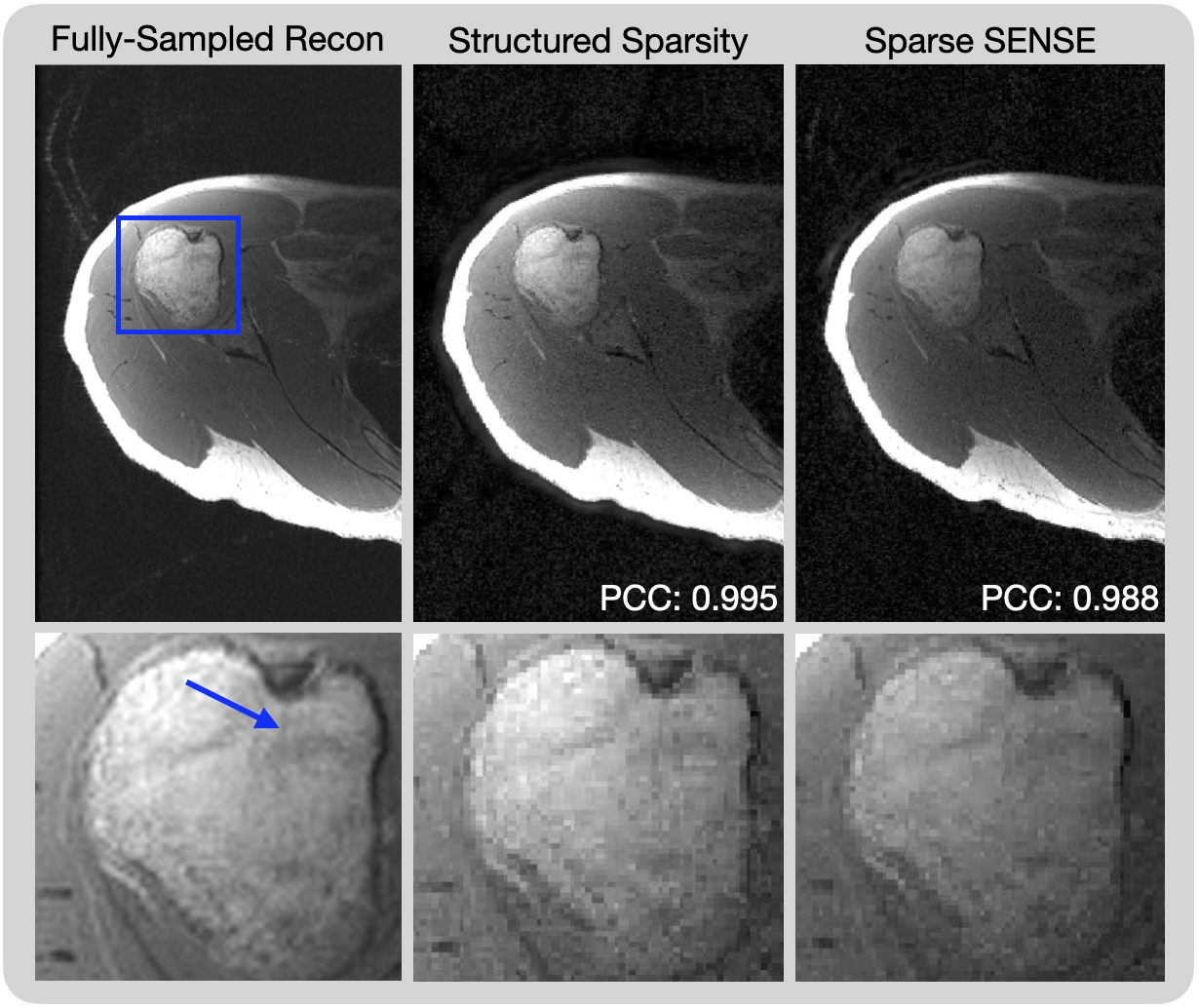}
    \caption{Reconstructions of an axial slice of a shoulder for compressed sensing with structured sparsity and Sparse SENSE.  The data collected had an acceleration factor of over $8$ (a sampling burden of $12\%$).  The top row shows the image reconstructions; the bottom row shows the region enclosed in the blue box enlarged for improved understanding of the details.  The blue arrow indicates a detail in the fully sampled image than can be more easily seen in the reconstruction with structured sparsity than it can in the reconstruction with compressed sensing alone.}
    \label{fig:shoulderResult}
\end{figure}

Figure \ref{fig:shoulderResult} shows a similar comparison between the fully-sampled reconstruction, are reconstructions from $12\%$ of the fully-sampled data using Sparse SENSE, and parallel imaging with compressed sensing using structured sparsity for data of an axial slice of a shoulder collected with a 16-channel shoulder array.  The top and bottom rows show the full image and an enlarged region, respectively.  As with the ankle and shoulder, compressed sensing with structured sparsity attains a better PCC than compressed sensing alone (PCC=$0.995$ and PCC=$0.988$, respectively).  The blue arrow indicates a detail that can be seen when structured sparsity is used but cannot be seen without it.

\begin{figure}
    \centering
    \includegraphics[width=0.9\linewidth]{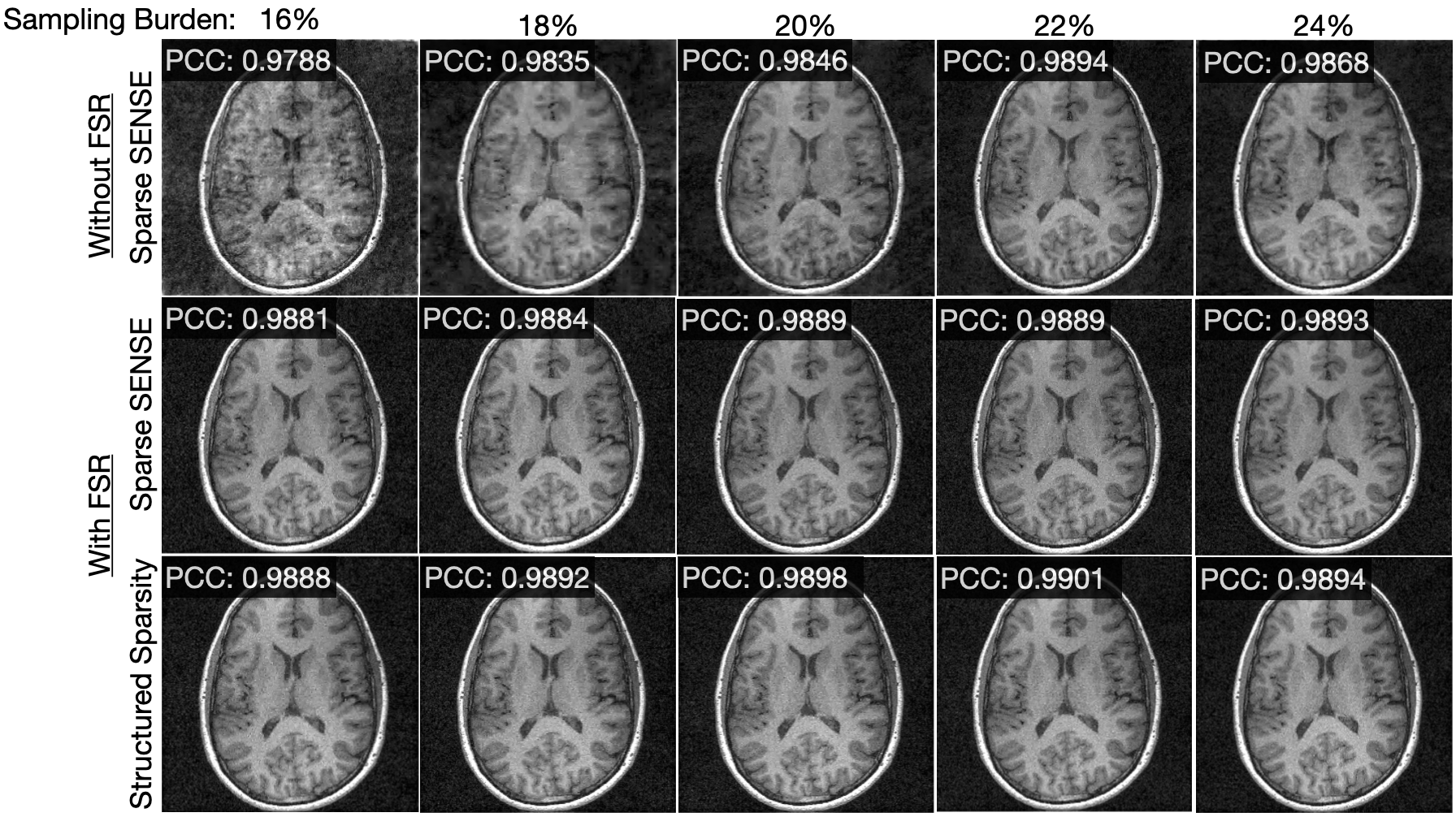}
    \caption{Reconstructions of an axial slice of a brain. For all sampling burdens, the PCC with structured sparsity is the highest.}
    \label{fig:pccRecons}
\end{figure}

Figure \ref{fig:pccRecons} shows reconstructions for an axial slice of a brain for a variety of different sampling burdens using a sparsifying transformation comprised of wavelets and curvelets.  In all cases, compressed sensing with structured sparsity outperforms Sparse SENSE.  Note that compressed sensing with structured sparsity achieves a PCC of $0.9892$ with a sampling burden of $18\%$, which is about what Sparse SENSE with the fully-sampled region achieves with a sampling burden of $24\%$.  This indicates that one can accelerate the MRI scan by an additional $25\%$ and achieve comparable or better image quality when taking advantage of structured sparsity.

\section{Conclusion}
\label{sec:conclusion}

When compressed sensing with structured sparsity is combined with parallel imaging, it achieves improved image quality over Sparse SENSE (which is compressed sensing and parallel imaging without structured sparsity).  The vast majority of the benefit is due to a sampling pattern that includes a fully-sampled region that centered on the $0$ frequency that satisfies the Nyquist-Shannon sampling theorem for the low-frequency bins of the wavelet and curvelet sparsifying transformations.  There is a small additional benefit by modifying the optimization problem to take the structured sparsity into account, due to the increased sparsity of the resulting optimization variable.

This manuscript presents compressed sensing with structured sparsity in the context of a model-based reconstruction \cite{fessler2010model}.  Compressed sensing with structured sparsity could also be integrated into parallel imaging based on linear predictability \cite{haldar2020linear}, such as SPIRiT \cite{murphy2012fast}, ESPIRiT \cite{uecker2014espirit}, or P-LORAKS \cite{haldar2016p}.  We leave this pursuit as future work.

\section*{Compliance with Ethical Standards}
All procedures performed in studies involving human participants were in accordance with the ethical standards of the institutional and/or national research committee and with the 1964 Helsinki declaration and its later amendments or comparable ethical standards.
MR data of humans was gathered with Institutional Review Board (IRB) approval and Health Insurance Portability and Accountability Act (HIPAA) compliance.  Informed consent was obtained from all individual participants included in the study.




\end{document}